\documentstyle[12pt]{article}
\begin{document}
\begin{titlepage}
\hoffset=.5truecm
\voffset=-2truecm

\centering

\null
\vskip 1truecm
{\normalsize \sf \bf International Atomic Energy Agency \\
and \\
United Nations Educational,Scientific and Cultural Organization\\}
\vskip 1truecm
{\huge \bf
INTERNATIONAL CENTRE \\
FOR \\
THEORETICAL PHYSICS\\}
\vskip 3truecm

{\LARGE\bf
Search For Dibaryonic De-Excitations  \\
In Relativistic Nuclear Reactions}\\
\vskip 1truecm

{\large \bf
{\bf C.Besliu
\\}
\normalsize University of Bucharest,P.O.Box MG-6,Bucharest,
{\bf Romania \\ }
\medskip
{\large \bf V.Popa,L.Popa\\}
\normalsize Institute for Space Sciences,P.O.Box MG - 6,Bucharest,
                               {\bf Romania}\\
\medskip
{\normalsize and}\\
\medskip
{\large \bf
V.Topor Pop\\}
\normalsize International Centre for Theoretical Physics,Trieste 34100,
{\bf Italy \\}
{\normalsize and\\}
\normalsize Institute for Space Sciences,P.O.Box MG - 6,Bucharest,
{\bf Romania\\}
{\it Revised version 20 July 1993\\}
\vskip 2truecm
{\it Work submitted to Journal of Physics G Particle and Nuclear
Physics}
}
\vskip 8truecm

\end{titlepage}

\hoffset = -1truecm
\voffset = -2truecm

\title{\bf Search For Dibaryonic De-Excitations\\
In Relativistic Nuclear Reactions}
\author{
{\bf
 C.Besliu
}\\
\normalsize University of Bucharest,P.O.Box MG-6,Bucharest,
{\bf Romania}\\
{\bf V.Popa,L.Popa}\\
\normalsize Institute for Space Sciences,P.O.Box MG - 6,Bucharest,
{\bf Romania}\thanks{E-MAIL POPAL@ROIFA.BITNET(EARN)} \\
{\normalsize and}\\
{\bf V.Topor Pop}\\
\normalsize International Centre for Theoretical Physics,Trieste 34100,
{\bf Italy}\\
{\normalsize and}\\
\normalsize Institute for Space Sciences,P.O.Box MG - 6,Bucharest,
{\bf Romania}\thanks{E-MAIL TOPOR@ROIFA.BITNET(EARN)}\\
{\it Revised version 20 July 1993}}
\date{20 July 1993}
\newpage

\maketitle

\begin{abstract}

Some odd characteristics are observed in the single particle
distributions obtained from $ He + Li $ interactions at
$ 4.5 AGeV/c $ momenta  which are explained as the manifestation
of a new mechanism of strangeness production via dibaryonic
de-excitations. A signature of the formation of hadronic and
baryonic clusters is also reported.The di-- pionic signals of the
dibaryonic orbital de-- excitations are analyzed in the frame of
the MIT - bag Model and a Monte Carlo simulation.The role
played by the dibaryonic resonances in relativistic nuclear
collisions could be a significant one.
\end{abstract}

\newpage



\section{Introduction}

  In the last years a wealth of data have accumulated concerning
relativistic nuclear interactions . The still growing interest
on this branch of physics has its roots in the hope to obtain
by the means of the relativistic nuclear physics information on
a hypothetical state of matter - the quark gluon plasma ,as well as
on other genuine properties of the compressed hadronic matter.
For reviews see references \cite{kn:koch},\cite{kn:carr},

\cite{kn:satz},\cite{kn:baym},\cite{kn:gut1},\cite{kn:tan1},

\cite{kn:sal},\cite{kn:gut2},\cite{kn:schuk}, \cite{kn:bamb}.

In this paper we discuss some unusual results obtained in the frame
of the SKM - 200  Collaboration from JINR - Dubna concerning the
yield of negative charged particles produced in the $He + Li$ interaction
at  \( 4.5 AGeV/c \). Despite of the much higher energies available today,
the few GeV sector is still interesting mostly because it offers the
perfect frame to perform studies on the properties of the exotic multi -
quark states, as those recently published on the production of non-strange
dibaryons in neutron - proton interactions at $ 5.1 GeV/c $\cite{kn:bes1},
 \cite{kn:bes2} . The theoretical description of those new states
\cite{kn:bes2} and its non - trivial consequences that should manifest when
they are produced inside a compressed nuclear matter as predicted in papers
\cite{kn:bes3},\cite{kn:bes4} are the motivation of this work.

\section{The Experimental Data}

  The results that we are going to analyze in the following are
originating from 3857 central collision events between a He beam
of  laboratory momentum $ 4.5 AGeV/c$ on a Li target inside the
2m $ SKM - 200 $ streamer chamber produced at the JINR - Dubna
synchrophazotron and represent a subset of the $ SKM - 200$
Collaboration data ,which cover a broad range of nuclear beams and
targets.The general experimental features are discussed in papers
\cite{kn:aksi},\cite{kn:anik},\cite{kn:abdu}. Some particular
results concerning the characteristics of the
$\pi^{-}$ production in O + Pb ,O + Ne
and in the bulk of He induced reaction are already presented in
references \cite{kn:bes5},\cite{kn:bes6},\cite{kn:bes7},\cite{kn:bes8}.
The experimental data processing procedure was typical for streamer
chamber experiments : as ionization measurements were not available,
the only particles that could be positively identified were the
negative pions.All tracks presenting secondary interactions with the
gas filling the chamber or signs of decay during the flight inside the
sensitive volume of the detector were removed in the first stages of the
scanning.In those conditions the averaged pionic multiplicity for
central $He + Li$ reaction is found to be \( <n_{\pi}^{-}> =1.757\pm .157 \).
We note that a simulation Monte Carlo HIJING code generator\cite{kn:gy1},
\cite{kn:gy2} applied in a Dual Parton Model variant give a value of
1.65  for pion multiplicities (we generate 50000 events)\cite{kn:top1}.
In paper of \cite{kn:anik} it is argued that the admixture of
$K^{-}$ and $\Sigma^{-}$ that escaped the scanning filtering is less
than $ 0.5 \% $.

An experimental details that is significant for the discussion in the
next sections is the position of the target inside the fiducial volume
of the chamber. The fact that the target is situated upstream respective
to the incident beam causes some loss in the detector selection power in
the backward reaction hemisphere.Figure 1 depicts a very schematic picture
of the chamber (drown at scale) and indicates also the projections in
the upper plane of the spectrometer of the three photographic cameras
(actually they are situated at 211.8 cm above that plane).Their "excentric"
positions are determined by the presence of the magnetic coils above the
chamber.

\section{Pecularities of Some Kinematical Single  Particle Distributions}

Most of the analysis published by the $ SKM - 200 $ Collaboration are
based on the study of the multiplicity of secondaries with respect to
the beam - target combinations and different degrees of centrality.
In the following we shall restrict our selves in the discussion of
some general kinematical distributions.

Figure 2 presents the angular distribution (in the laboratory frame)
of the negative particles ,which are , in concordance with the
statements in the second section, supposed to be $\pi^{-}$ mesons.

An unexpected high production of particles in the backward hemisphere
(azimuthal angles higher than $137^{o}$ thus in the cone represented in
Figure 1 ) is striking.The same conclusion can be drown from the rapidity
distribution depicted in Figure 3 .While in the rapidity range
$ y > -0.7 $ the shape of the distribution is quite a normal one ,
at $ y < -0.7 $ the data manifest some abnormal behavior.

Both the angular and rapidity distributions show that mixed negative
mesons of different origin are present in the experimental data .
Furthermore ,the anomalously high absolute values of the rapidity of
this component (the rapidity being computed as those particles being
pions) could be reduced if assuming the corresponding particles being
more massive.The cut - off of the contributions of this puzzling
component towards forward directions suggests that the filtering of
negative pions through the scanning procedure is not operational in
the upstream direction,namely in that direction characterized by
significantly shorter distances available inside the fiducial volume
of the chamber.Unfortunatelly such an interpretation is not offering a
satisfactory explanation of the sharpness of the observed cut - off.
We shall return to this problem in section 4  of this paper.

The most reasonable guess leading to an acceptable understanding of
the manifested anomalies is to assume that an important amount of
$ K^{-} $ mesons is produced during the interaction and that their
decay escapes from detection within the active volume of the chamber
when produced backwards.

If this assumption is true , then this kaonic component should be
produced ( according the conventional picture of high energy nuclear
interaction )in an earlier stage of the hadronization,namely at higher
temperature than the pionic component.Information about the production
temperature could be extracted from the transverse momentum
distributions.This distributions present  also the advantage to be
independent on the mass values of the particles involved ,as momenta
are direct observable in the experiment due to the magnetic deflection of
the charged secondaries inside the chamber .Figure 4 depicts the
$ p_{\bot} $ distribution for all the negative particles (the circles),
the particles with \( y_{lab} \geq -0.7 \) which could be considered as
$\pi^{-}$ mesons (the squares) and for those with \( y_{lab} < -0.7 \),
namely those we suppose to be $ K^{-} $ mesons (the rhomboids).The
solid lines in Figure 4 are obtained by fitting the descending parts of
the $ p_{\bot}$ spectra with exponential functions:

\vskip 0.3cm

 \[ \frac{\partial N}{\partial p_{\bot}} \propto exp[-\beta \cdot
p_{\bot}]   \]

\vskip 0.3cm

where the fitting parameter $\beta $ should be in  inverse
 proportionality with the source temperature . The results of
 those fits are listed in  Table 1 .

The conclusion that could be extracted from those results is that the
anomalous negative particle component observed in the $He + Li$
central collisions at $4.5 AGeV/c$ originates from an earlier
stage of the interaction - as characterized by a higher temperature
 than the usual $\pi^{-}$ component .This observation is supporting
our claim that strong $K^{-}$ production manifests  in the investigated
reaction.
\vskip 0.3cm

{\bf Table1\/} The parameters of the exponential functions describing
     the $ p_{\bot}$ spectra
\vskip 0.3cm

\begin{center}

\begin{tabular}{|c|c|c|}
\hline \hline
{\bf  Selection} & \( \beta (GeV/c)^{-1} \) & \( \chi^{2}/{N.D.F.} \)  \\
\hline
{\bf   all}       & $1.61 \pm  .04 $  & $1.29$ \\
\hline
{\bf $y\geq -0.7$} & $2.49 \pm .03$ & $1.66$ \\
\hline
{\bf $y < -0.7$}   & $1.02 \pm .03$ & $1.41$ \\
\hline \hline
\end{tabular}

\end{center}
\vskip 0.3cm

The high kaonic multiplicity \( <K^{-}> = 0.712 \pm 0.008 \) should be
analyzed in the context of the averaged number of participants
nucleons in the investigated reaction. Taking into account that
in the $He + Li$ reactions equal numbers of protons and neutrons
are involved,one may estimate this number using the relation :
\vskip 0.3cm
\begin{equation}
   <N> = {2}\times (<Q -2\times n^{-}-n_{s}>)
\end{equation}
\vskip 0.3cm

where Q represents the charged particle multiplicity ,$n^{-}$ the
number of negative particle/event and $n_{s}$ the number of
stripping particles/event. The value of $<N>$ (averaged for all
the 3857 investigated collisions) is \( <N>=5.854 \pm 0.393 \) ,
while the average number of participant protons is found to be
  \( <p> = 2.972 \pm 0.197 \) .Considering that the production of kaons
is izospin symmetric ,it would follow that the K/p ratio in
$He + Li$ collisions at $4.5 AGeV/c$ is \( (95.8 \pm 6.9)\% \),
about 20 times greater than in pp collisions $ (2 -5) \% $ .

Such unexpectedly large kaonic production asks for an untrivial
explanation. It may be argued that this ratio is even higher ,
due to the elimination in the initial scanning of the kaons
produced in the forward direction but the production mechanism
discussed in the next section rules out the possibility of such
an extrapolation.

\section{Strangeness Production Trough Dibaryonic
 De -- excitations}

In paper of \cite{kn:bes2},a model for non-strange dibaryons is
developed starting from the conventional M.I.T bag model
calculations and its predictions are shown to be in good
agreement with the experimental data ,mostly those obtained at
J.I.N.R. - Dubna from the irradiation of the 1 m hydrogen
bubble chamber with a quasi-monochromatic beam of neutrons.
The originality of that description resides in assuming the
dibaryonic system composed of a diquark ( a bound but unconfined
two quark aggregate ) and a cluster of four nearly free quarks.
This assumption enables the six quark bag to acquire some degree
of stability even in the $l=0$ orbital momentum state . The
$l \geq 1$ states are found to have masses allowing their direct
decay into ($NN n\pi$) channels with n=1,2 and even 3.Experimental
candidates for $n = 1$ are also identified in the $np$ experiment
and are analyzed in the same paper.The experimental data
published by \cite{kn:bes2} allow us to estimate the importance
of the dibaryonic production in nucleonic interactions at energies
close to those of the $He + Li$ experiment .At $p_{inc}=5.1 GeV/c$
the sum of the total cross sections of the three five -prong
channels investigated is \( \sigma_{T} = 1520 \pm 110 \mu b\, \),while
the sum of the reported dibaryonic production cross sections is
\( \sigma_{D}=601^{+30}_{-61} \mu b\, \),thus representing
a percentage of  \( 39.5 \% _{-21.1\%}^{+24.2\%} \) of the total cross
section (the asymmetry of the errors originates as shown earlier
 \cite{kn:bes2},from the interference between the fit errors
 affecting the weights of the dibaryonic peaks and those of the
fitted widths).The sum of the production cross sections of the $l = 1$
 dibaryonic candidates exceeds about four times that corresponding
 to the $l = 0$ dibaryons  \( 480_{-51}^{+218} \mu b\, \) versus
respectively \( 120_{-10}^{+83} \mu b\, \) .It must be noticed that
in he conditions of the $np$ experiment ,
 there are some forbidden dibaryonic izospin states ,
 thus the  results above are slightly under - estimated .
It must be also  stressed that in that experiment the contribution of
dibaryonic states with $l \geq 2$ was not taken into consideration,
but their occurance was signaled .

It was shown \cite{kn:bes3},\cite{kn:bes4} that the orbital de - excitations
are likely to produce within different dibaryonic states.The cascade
transitions \( (l = 2) \Rightarrow (l = 1) \Rightarrow (l = 0) \) create
a complicate structure of narrow maxima in the beginning of the
di-pionic effective mass spectrum , while the direct
\( (l = 2) \Rightarrow (l = 0) \) de - excitations with the emission of
a pair of pions are shown to lead to a significant bump in the $2\pi$
invariant mass spectra ,near the threshold , quite similar to those
reported in the \( M_ {\pi^{-}\pi^{-}} \) effective mass spectrum from
 $O + Pb $ central collisions at $4.5 AGeV/c$  \cite{kn:bes8}. This
similarity supports the hypothesis of significant dibaryonic
production in nucleus  nucleus collisions in few GeV energy region.

An other important consequence of the diquark -four-quark cluster
structure for the dibaryonic resonances ,also discussed in papers
of \cite{kn:bes3},\cite{kn:bes4} is the strong dependence of the dibaryonic
mass spectra and of the slopes of the corresponding Regge
trajectories on the density of the environmental nuclear matter.
It was shown that if the density of the production region enough
high ,the threshold for the production of kaons via dibaryonic
de - excitations is over passed ,thus opening supplemental channels
of strangeness production and leading to $S = -1 $ dibaryons in the
ground orbital state .

A typical diagram describing a $\Delta l = 1 $ dibaryonic
de-excitation with the emission of a kaon is presented in Figure 5.
The rearrangement of quarks (the s quark of the created pair is
substituted by a nonstrange quark from the four-quark dibaryonic
cluster) is quite mandatory ,as the available de-excitation energy
is not enough in order to allow the hadronization of the $(s,\bar s)$
pair as $ \eta $ meson . The resulting dibaryon should be observable
as a resonant $(N\Lambda)$ state , as predicted by \cite{kn:popa}
where a version of our dibaryonic model have been tested for the
description of the $(p\Lambda)$ and $(p\Lambda\pi)$ observed
candidates \cite{kn:shah} . The absence of the $ \bar u $ and $ \bar d $
antiquarks in the system rules out the production trough diagram
(Figure 5) of both the $K^{-}$ and $\bar {K}^{o}$ meson .

{}From the kinematical point of view ,the diagram (Figure 5)
represents a two body decay,in which only scalar particles
are involved.In such conditions the ground state dibaryon and
the emitted meson (kaon or ,in the case of the gluon decay into
a pair of nonstrange quarks a pion ) are emitted with opposite
momenta on a uniformly distributed direction with respect to the
reference frame of the decaying state .As initial dibaryons are
likely to be produced as a result of the interaction between the
beam nucleons with the target ones,they should (in the laboratory
reference system) move in the forward direction.Taking into
consideration the fact that the $l = 0$ dibaryon carries out the
most important fraction of the mass of the $l = 1$ one , it is
reasonable to suppose that the de-excitation mesons are
preferentially emitted in the backward hemisfere , so the kaon
production in this direction should not be extrapolated to the
forward region of the reaction .Those simple kinematical arguments
may offer a convenient explanation to the fact that the observed
anomalous backwards kaon production ( at least in which concerns
the positive unidentified kaons ) should not manifest itself with
the same intensity in the forward direction.

In order to appreaciate the importance of the diagram depicted in
Figure 5 to the total yield of straniety in our reaction,it is
significant to analyze it in balance with the corresponding
nonstrange diagrams which would lead to pion production.As pointed
out in papers \cite{kn:bes3},\cite{kn:bes4},the increase of the
dibaryonic Regge trajectories in the conditions of nuclear matter
densities reasonable to expect in collisions involving light nuclei ,
results in de - excitation energies close to the kaonic threshold.
We may then suppose that the diagrams without strangeness production
would preserve in our experimental conditions , the leading role.

The explanation of abundant $K^{-}$ production is to be found by a
different mechanism.As suggested by the $np$ experiment ,the
production of $l = 2$ dibaryonic states is also significant in
nucleon nucleon collisions at close energies.In the conditions of
collisions between relativistic nuclei , due mostly to the
collective interactions, the production of such excited multiquark
states could be enhanced.Those systems could de-excite to the
ground state by a succession of two $\Delta l = 1$ processes as
that described above (which would be less probable due to the
finite life of the dibaryonic resonances),or directly as suggested
by the diagram of the type presented in Figure 6.

The available energy for such a de-excitation process is two times
that available in the previous cases so the strangeness production
should be favour with respect to the nonstrange channels.

The first gluon exchanged between the four quark cluster and the
diquark is supposed to leave the dibaryonic system in the $l = 0$
state and thus should carry two units of kinetic momentum.If the
gluon decays into a pair of quarks $ (s,\bar s) $ those quarks will be
inhibited to hadronize due to their large relative orbital
momentum.The hadronization becomes possible after the exchange
of the second gluon between the two strange quarks that could
decay into a $(u,\bar u)$ or a $(d,\bar d)$ quark pair ,thus leading in
the final state to the creation of the $(K^{+},K^{-})$ or a
$(K^{o},\bar {K}^{o})$ pair , such a diagram being the only one that
could be responsible of the anomalous $K^{-}$ production observed
in our data .As the pair of mesons are simultaneously produced
with a certain delay after the dibaryonic de - excitation took
place ,the entire process could be regarded to be more likely a
two - particle decay than like a three - particle one.It follows
from this assumption that the kinematical distribution of the
kaons is like to be shifted towards the backward hemisphere ,
from the same reason as in the previous case.

The diagram from figure 6 offers also the possibility to make
semi-quantitative analysis of the conditions in which the
observed yield of $K^{-}$ mesons is a reasonable one.As the
experimental values of $<K^{-}>$ is about 0.7 and as the
probability to obtain a negative kaon from the investigated
process is $ 50 \% $,the requested averaged number of dibaryonic
$\Delta l = 2$ de - excitations in one $He + Li$ interaction
must be about 1.4 .From the dibaryonic production cross sections
measured in the $np$ experiment it follows that the probability
of the creation of such a resonances (only for the $l = 0$ and
the $l = 1$ ones ) is about $ 40 \% $ for each individual interaction.
Assuming that in $ He + Li$ reaction the production of $l = 2$
dibaryons has the same order of magnitude  and tacking into
account the averaged number of participant nucleons $<N>\cong 6$,
it follows that  the expected number of $l = 2 $ states should be
about 1.2 .As nuclear effects could actually enhance the excited
dibaryonic production,such a mechanism could thus lead to a
reasonable quantitative agreement with the experimental data .

We remark also that Jacob and Rafelski \cite{kn:jac} have pointed out
that in a baryon rich environment which also contains high density
of u and d quarks and antiquarks a higher relative abundance of
strange quarks should be favored .

This result confirms also the recent identification of hadronic
clusters in central collisions reported by El Naghy \cite {kn:nag}
by more complex analysis of the emission of the produced shower
particles and target fragments in $Si + Ag(Br)$ and $Mg + Ag(Br)$.
The observed back - to - back emission of these clusters reflects
the sideward flow of nuclear matter in the studied interactions.

\section{Di-Pionic Effective Mass Anomalies As Dibaryonic Signals}

Assuming the above interpretation of hard negative secondaries
measured in the backward hemisphere as being kaons produced
trough dibaryonic de - excitations , it follows that signatures
of those controversial resonances should also manifest. In
papers \cite{kn:bes3},\cite{kn:bes4}, it was speculated that the
dibaryonic de-excitations via pionic emission should lead to anomalous
maxima in the low mass region of the invariant mass spectra of
two pions .Such effects have been already reported , as observed
in connection with dibaryonic production by neutron - proton
interaction at a few GeV and could be a tentative explanation
to the so - called "ABC" effect \cite{kn:bes3}.Strong anomalies in
the \( (\pi^{-}\pi^{-}) \) and \( (\pi^{-}\pi^{-}\pi^{-}) \) invariant
mass spectra have been also reported in $O + Pb$ reaction at
$4.5 AGeV/c$  \cite{kn:bes1} . This last example belongs also to the
$SKM - 200$ Collaboration and it has been for the first time suggested
that such maxima could be related to the dibaryonic production .

In Figure 7 we present the effective mass spectrum for the pions
produced in the $He + Li$ collisions at $4.5 AGeV/c$ . The bin
width was chosen in concordance with the estimated experimental
resolution.The contribution of particles with $y \leq -0.7$ was
ommitted.It must be underlined that Figure 7 depicts only the
low mass region of the spectrum . The tail of the invariant
di - pionic mass distribution observed in our interaction expands
till about $ 2 GeVc^{-2}$ , but for the signals we search for only
the represented mass region is relevant .

In order to construct the background distribution , we have combined
pions originating from a different event from a set of 200
interactions,obtaining the upper dotted (the larger dots) in
Figure 7 .By fitting the spectra with only this contribution we
have obtained a value of the \( \chi^{2}/N.D.F. \) of 1.579 . The
inclusion of two components of the spectra simulated in papers
 \cite{kn:bes3},\cite{kn:bes4} representing the contribution of the
direct dibaryonic de - excitations with $\Delta l = 2 $ and of cascade
de - excitations from the predicted $l = 3$ states to $l = 0$
dibaryonic states by successive $\Delta l = 1$ transitions,
represented in Figure 7 by the lower (smaller dots) dotted lines,
resulted in a decrease of the \( \chi^{2}/N.D.F \) to the value of
0.93 . The solid line in Figure 7 indicates the best fit curve,
the contribution of the two kind of dibaryonic de-excitation being
$ 2.8 \% $ ( for the direct ones ) and $ 2.5 \% $ for step-by-step ones.
The fit was made considering the full invariant mass spectra(60beans)
which could explain the small values of the above contributions.
If we take into consideration the fact that the de-excitation
signals affect only the first ten beans of the spectrum,then the
improvement of the \( \chi^{2}/N.D.F \) estimator with about 0.5
indicates an integrated effect of approximatively five standard
deviations for those points.

The insufficient experimental statistics (due to the low pionic
multiplicities of the investigated reactions did not allow us to
introduce further hypothesis in order to improve the concordance
between the spectrum and the simulated curve,but we may claim
that the di-pionic signals of the dibaryonic orbital
de - excitations were identified in our data .

The observed excess of pion pair production with low values  of
the invariant mass is equivalent to an enhancement of the
production of pions with low relative momenta .This effect could
be related with the observed angular correlation maxima at low
relative angles in $p + Au$ collisions at 4.9 ,60 and 200 GeV ,
as reported in paper \cite{kn:kamp}.The explanation proposed in
that article is based on the geometric properties of hadron--
nucleus interaction at non - - vanishing values of the impact
parameter , but such an argument seems less convincing in the
case of nearly symmetric nucleus-nucleus collisions.

\section{Conclusions}

The unusual behavior of some uniparticle distributions observed in
the $He + Li$ interaction at $4.5 AGeV/c$ is explained as the
effects of an increased production of $K^{-}$ mesons via dibaryonic
de - excitation .Di-pionic invariant mass spectra are found to be
in agreement with the expected signals of dibaryonic orbital
de - excitations , thus supporting our interpretation of data .

The role played by the dibaryonic resonances in the relativistic
nuclear collisions could be a significant one. The experimental
data analyzed support the previous description of those exotic
states as diquark - four quark cluster bags ,as well as the
predicted consequences concerning the behavior of such systems
if produced inside compressed nuclear matter. The production
of dibaryons during the high energy nuclear collisions could be
responsible on the lack of observing till so far the creation
of quark - gluon plasma in such processes .In fact , the
apparition in the compressed nuclear matter of a bosonic phase,
made of dibaryonic resonances,is expected to inhibit the
deconfinement of quarks as supposed by the conventional picture
of the relativistic nuclear physics .

We claim that the wealth of pion and kaon data accumulated in
experiments at higher energies and multiplicities (CERN and BNL)
could ofer,if appropriately analyzed ,very relevant insights on
the role played by an intermediate bosonic phase in relativistic
nuclear collisions.

\section{Acknowledgements}

We would like to thank  our colleagues and  workers M.Anikina,
A.Goloskhvastov, K.Iovcev, S.Khorozov, E.Kuznetzova, J.Lukstins,
E.Okonov, G.Vardenga, M.Gazdzici, E.Skrypczak and R.Swed and all
other members of the $SKM - 200$ Collaboration for their essential
contribution .

One of us (V.Topor Pop) express his gratitude to Professors
Giorgio Belletini ,Hans Gutbrod and Johann Rafelski for their
kind invitation and for the NATO grant which allowed him to
participate in an very well organized NATO -ASI school.

I would like also to thank to Professor Miklos Gyulassy and
Dr.Xin-Nian Wang for providing the Monte Carlo program
HIJING and many helpful communications.

Also express his gratitude to Professor Abdus Salam and to
International Atomic Energy Agency and UNESCO for hospitality
at the International Centre for Theoretical Physics ,Trieste
where this paper was completed .

Special thanks are due to the referees of the first version
for their pertinent observations which improved  this paper.

\newpage

\centering

\begin{figure}
\vskip 0.3cm
\label{Figure1}
\caption{Schematic picture of the streamer chamber}
\vskip 0.3cm
\label{Figure2}
\caption{The azimuthal distribution of the negative secondaries
in He + Li central collisions at 4.5 AGeV/c}
\vskip 0.3cm
\label{Figure3}
\caption{The rapidity distribution of the negative secondaries
in $ He + Li $ central collisions at $ 4.5 AGeV/c $}
\vskip 0.3cm
\label{Figure4}
\caption{The $p_{\bot}$ distributions for the negative secondaries
in $ He + Li$ collisions}
\vskip 0.3cm
\label{Figure5}
\caption{Diagram describing the proposed $\Delta l = 1$
dibaryonic de - excitations }
\vskip 0.3cm
\label{Figure6}
\caption{Diagram describing the proposed $\Delta l = 2$
dibaryonic de - excitations }
\vskip 0.3cm
\label{Figure7}
\caption{The \( ( \pi^{-}\pi^{-}) \) invariant mass spectrum in He + Li
central collisions at 4.5 AGeV/c.The curves are theoretical results
obtained from different dibaryonic de - excitation (see the text for
explanations)}
\vskip 0.3cm
\end{figure}

\end{document}